\documentclass[twocolumn,usenatbib]{mnras}

\usepackage[T1]{fontenc}
\usepackage{ae,aecompl}

\usepackage{graphicx}	
\usepackage{amsmath}	
\usepackage{amssymb}	

\title[Local fragmentation of thin ...]{Local fragmentation of thin disks in Eddington-inspired gravity}

\author[Roshan et al.]{
Mahmood Roshan,$^{1}$\thanks{E-mail: mroshan@um.ac.ir}
Ali Kazemi,$^{1}$
Ivan De Martino $^{2}$
\\
$^{1}$Department of Physics, Ferdowsi University of Mashhad, P.O. Box 1436, Mashhad, Iran\\
$^{2}$Department of Theoretical Physics and History of Science, University of the Basque Country UPV/EHU,\\
 Faculty of Science and Technology, Barrio Sarriena s/n, 48940 Leioa, Spain
}

\date{Accepted XXX. Received YYY; in original form ZZZ}

\pubyear{2018}

\begin{document}
\label{firstpage}
\pagerange{\pageref{firstpage}--\pageref{lastpage}}
\maketitle

\begin{abstract}
We find the generalized version of the Toomre's criterion for the stability of a rotating thin disk in the context of Eddington inspired Born-Infeld (EiBI) gravity which possesses one free parameter $\chi$. To do so we use the weak field limit of the theory and find the dispersion relation for the propagation of matter density waves on the surface of a self-gravitating and differentially rotating disk. Finally we find a new version of Toomre's stability criterion for thin disks. We show that EiBI gravity with negative $\chi$ destabilizes all the rotating thin disks. On the other hand EiBI with positive $\chi$ substantially can suppress the local fragmentation, and has stabilizing effects against axi-symmetric perturbations. More specifically, we show that only an annulus remains unstable on the surface of the disk. The width of the annulus directly depends on the magnitude of $\chi$.
\end{abstract}




\section{Introduction}

General Relativity (GR) needs resorting to Dark Matter and Dark Energy in order to explain the dynamics of
self-gravitating systems at galactic and extragalactic scales, and the evolution of Universe as a whole \citep{Planck16_13,Feng2010}. 
Although many dark candidates have been proposed \citep{Bertone2005,Capolupo2010,demartino2017b,Capolupo2017}, the fundamental nature of these components is still unknown, and the fact that their introduction can not be avoided, has often been interpreted as breakdown of GR. One path to overcame such emerging shortcomings seems to point to generalize the gravitational action. It has been demonstrated that modified theories of gravity have the capability to explain both the dynamics of self-gravitating systems and the cosmological evolution of the space-time
\citep{Nojiri2011,PhysRept,Annalen,Clifton2012,demartino2014,idm2015,demartino2016,Cai2016,Beltran2018, Nojiri2017}. Alternatives to the Einstein-Hilbert action should, anyway, 
be considered. In fact, the singularity theorems of GR tell us that the presence of space-time singularities inside black holes and in the early universe are unavoidable \citep{Hawking1973}, and that the widespread lore in the field states that this is due to the fact that GR breaks down near such singularities, due to the fact that the quantum effects of gravity cannot be longer neglected there. One way around has been found in the context of Eddington-inspired Born-Infeld (EiBI) gravity  \citep{Banados2010}. As another example for this case see Energy-Momentum-Squared Gravity (EMSG) proposed in \cite{Roshan2016}.

\citet{Banados2010} proposed a new class of theories inspired to the Born-Infeld non-linear electrodynamics. In such a way, EiBI is capable to avoid the existence of some of the singularities plaguing GR \citep{Delsate2012} introducing new couplings between the gravitational field and the matter fields. As an example, the formation of singularity is prevented in cosmology \citep{Banados2010} as well as in the context of the gravitational collapse of compact objects \citep{Pani2011,pani}. However, we should note that it could be the source of some anomalies in compact stars \citep{Pani2012b, Sham2013} and we should mention that \citet{Bouhmadi2014} have shown that EiBI is not capable to avoid the Big Rip singularity. Nevertheless, EiBI represents a very interesting framework since it reduces to GR in the vacuum while its effects arise  in dense matter environments, such as neutron stars, where GR is experimentally not well probed. Besides, 
any modified theories of gravity must be able to reproduce the general relativistic results in low energy environments. Indeed, the analysis of the Newtonian limit has often been used to retain or to rule out modified theories of gravity (see for example \citet{Capozziello2012, Farinelli2014,Roshan2014,Sharif2014a,Sharif2014b,Hendi2015, Noureen2015, Sharif2015,  Vainio2016, idm2017a}).

Although EiBI gravity has been constrained using a wide range of probes, such as solar observations \citep{Casanellas2012}, compact objects \citep{Pani2011,pani,Avelino2012b,Sham2012,Sham2013,Harko2013,Sotami2014}, and the cosmological evolution of the space-time \citep{Avelino2012a,Avelino2012b,deFelice2012,Bouhmadi2015,Bouhmadi2017}, its consequences on low energy systems, such as the dynamics and the evolution of self-gravitating clouds,  are poorly explored. \citet{idm2017a} have investigated the stability criteria of self-gravitating systems pointing out that the Jeans instability is modified only in high densities environments, such as neutron stars, while EiBI gravity's effects become negligible in star formation regions. However, no studies of stability of  differentially rotating disks have been done yet.

In  Newtonian dynamics, the stability criteria to all local axisymmetric perturbations was firstly investigated by \citet{Toomre1964} who defined
a dimensionless quantity $Q$, for fluid and stellar disks, to characterize the stability condition, $Q>1$, of the system. 

On the other hand, in the cases where $Q < 1$,  the gravitational force in overdense regions overcomes the thermal pressure and rotation that can not prevent the collapse (for a review of the subject we refer the reader to see \citealp{binney}). Generalizing the  Toomre's  criterion for the local  stability  in  the  framework  of EiBI gravity could potentially be very useful to describe the dynamics of accretion disks around massive object.
On the other hand, the new Toomre's criterion can, in principle, lead to new and testable results. For example, the local stability of a system can be totally different in the context of different gravity theories. As we will show in this paper, rotating thin disks which are stable in standard gravity, are unstable in EiBI gravity with $\chi<0$. This is a testable result, and observation may help to check the validity of the theory.
 For similar studies in the framework of other modified theories of gravity see \cite{Roshan2015a} and \cite{Roshan2015b} for scalar-tensor-vector theory and $f(R)$ gravity respectively.

The paper is organized as follows.  In section \ref{EiBI_Gravity} EiBI gravity and its weak field approximation is briefly reviewed. In section \ref{DR}, we derive the dispersion relation for a self-gravitating fluid disk in the context of EiBI gravity using the first order perturbed equations.  Furthermore, using the modified dispersion relation, we derive the local stability criterion and generalize the Toomre's stability criterion to be valid in EiBI gravity. We show that this theory leads to interesting deviations from standard gravity at high density regime. Furthermore in section \ref{expon} we apply the results to an exponential thin disk model. Using this toy model we study the possible differences between EiBI and Newtonian gravity. Also in section \ref{applications}, we we apply our results to disks around Hyper Massive Neutron stars (HMNS). Finally in section \ref{conc} we discuss and summarize our main results.

\section{A brief introduction to EiBI gravity}\label{EiBI_Gravity}
In this section we will briefly summarize the EiBI theory of gravity. The starting point of any
relativistic theory of gravity is the action which, in the case of EiBI gravity, takes the form:
\begin{equation}\label{eibi_action}
 S=\frac{2}{\chi}\int d^4x\biggl(\sqrt{|g_{\mu\nu}+\chi R_{\mu\nu}|}-\lambda\sqrt{-g}\biggr)+S_{matter}[g,\phi_M].
\end{equation}
Here  $\phi_M$ is a generic matter field, $R_{\mu\nu}$ represents the symmetric part  of the Ricci tensor built from the affine connection $\Gamma$. The latter ones  are derived from an auxiliary metric denoted $q_{\mu\nu}\equiv g_{\mu\nu}+\chi R_{\mu\nu} $ satisfying the Eddington  field  equations, 
$\sqrt{|g_{\mu\nu}+\chi R_{\mu\nu}|}$ is the Born-Infeld like  structure, $\chi$ is a new EiBI parameter\footnote{It is customary in literature to indicates the EiBI parameter with $\kappa$. However, we decided to change the notation of the EiBI parameter to preserve the one of the epicyclic frequency, also indicated with the same greek letter, needed to describe rotating disks (see Eq. \eqref{epip}).} having the dimensions of length square, $g$ and $q$ refer to the determinants of the metric $g_{\mu\nu}$ and $q_{\mu\nu}$, and finally,
$\lambda$ is a dimensionless constant linked to the cosmological constant $\Lambda$ as: $\Lambda=\frac{\lambda -1}{\chi}$, for more details see \citep{Deser1998, Banados2010, Beltran2018}. Therefore, asymptotically flat solutions corresponding to $\Lambda=0$ are obtained for $\lambda=1$. Finally, the field equations are built in the Palatini approach. Although there are some classes of modified gravity models, belonging to the Lovelock family (see for details \citet{Borunda2008}), for which the metric and Palatini approach are equivalent, this is not guaranteed in general. Nevertheless, it is more convenient than the metric one that requires additional terms in the gravitational action to avoid ghost solutions \citep{Deser1998,Vollick2004}.  
Thus, varying the action (\ref{eibi_action}) independently with  respect to  the metric and the connections, one obtains the following set of field equations:
\begin{align}
\sqrt{q}q^{\mu\nu} & =\sqrt{g}\left[(1+\chi\Lambda)g^{\mu\nu}-8\pi\chi T^{\mu\nu}\right]\,, \\[3mm]
 0 & =\tilde{\nabla}_\sigma [\sqrt{q}q^{(\mu\nu)}] -\tilde{\nabla}_\gamma  [\sqrt{q}q^{\gamma(\mu}]\delta_\sigma^{\nu)}.
\end{align}
These are a set of second order differential equations in the metric tensor $g_{\mu\nu}$ containing also second derivatives
of the stress-energy tensor. Studying the local effects of the matter field in the non relativistic  weak field limit, the gravitational potential $\Phi ({\bf r},t)$ and the matter density $\rho ({\bf r},t)$ are linked by the following modified Poisson equation:
\begin{equation}\label{modPoiss}
 {{\nabla }^2}\Phi ({\bf r},t) = 4\pi G\rho ({\bf r},t) + \frac{\chi}{4}{{\nabla }^2}\rho ({\bf r},t)\,.
\end{equation}
Here the term $4\pi G\rho ({\bf r},t)$ is the standard Newtonian contribution while the second one, $\frac{\chi}{4}{{\nabla }^2}\rho ({\bf r},t)$, gives rise to a new source of gravity. Since this new term depends on the derivatives of the matter density, deviations from Newtonian gravity will arise in such environments where those terms are not negligible. Using the Poisson Equation (\ref{modPoiss}), the additional term induces a correction to the gravitational force which can be interpreted as a gradient of the effective pressure $p_{\text{eff}}=\chi \rho^2/3$. Therefore, depending on the sign of the free parameter, EiBI gravity can prevent the gravitational collapse. An important consequence of the Eq.	(\ref{poiss}) is the emergence of a new scale length, $k^2_\text{EiBI}=\frac{16\pi G}{|\chi|}$, a part of the classical Jeans scale $k_J$ \citep{Avelino2012c, idm2017a}. By comparing  the
electromagnetic and gravitational interactions inside atomic nuclei \citet{Avelino2012c} constrained
$ |\chi| < 10^{-3} \, {\rm kg^{-1} \, m^5 \, s^{-2}}$. Otherwise, nuclei 
would not exist. 

Now the aim is to use the hydrodynamics equations in the Newtonian limit to obtain a criteria for the stability of a self-gravitating fluid disk trough a modified Toomre's criteria.  

\section{DISPERSION RELATION FOR A SELF-GRAVITATING FLUID DISK IN EiBI gravity}\label{DR}

The dynamics of a  noninteracting pressureless self-gravitating system is governed,  in nonrelativistic limit, by the Euler and continuity equations that are modified by means of the modified Poisson's equation \eqref{modPoiss} (see for more details \citet{Pani2011, pani, Avelino2012b, Beltran2018}). 
Therefore the governing dynamical equations for a fluid system in EiBI are given by
\begin{equation}
\frac{\partial \rho}{\partial t}+\nabla\cdot (\rho \mathbf{v})=0\,,
\label{cont}
\end{equation}
\begin{equation}
\frac{\partial \mathbf{v}}{\partial t}+(\mathbf{v}\cdot \nabla)\mathbf{v}=-\frac{\nabla p}{\rho}-\nabla \Phi\,,
\label{euler}
\end{equation}
\begin{equation}
\nabla^2 \Phi=4 \pi G \rho +\frac{\chi}{4}\nabla^2 \rho\,,
\label{poiss}
\end{equation}
where $\mathbf{v}$ is the velocity field of the fluid, and $p$ and $\rho$ are the pressure and matter density respectively. Naturally, we need an equation of state in order to make a complete set of differential equations for studying the dynamics of the fluid.

In order to find the dispersion relation for the perturbations in the non-rotating cylindrical coordinate system $(r,\varphi,z)$, let us first linearize the equations (\ref{cont})-(\ref{poiss}). To do so we perturb the physical quantities, collectively shown as $\mathcal{Q}=\mathcal{Q}_0+\mathcal{Q}_1$, where the subscripts "$0$" and "$1$" stand for the background value and the first order perturbations one, respectively. In this case after some straightforward manipulations, one can show that the linearized version of continuity and Euler equations are
\begin{equation}
\frac{\partial \Sigma_{1}}{\partial t}+\frac{1}{r}\frac{\partial}{\partial r}\left(\Sigma_{0} r v_{r1}\right)+\Omega \frac{\partial \Sigma_{1}}{\partial \varphi}+\frac{\Sigma_{0}}{r} \frac{\partial v_{\varphi 1}}{\partial \varphi}=0\,,
\label{cont1}
\end{equation}
\begin{equation}
\frac{\partial v_{r1}}{\partial t}+\Omega \frac{\partial v_{r1}}{\partial \varphi}-2\Omega v_{\varphi 1}=-\frac{\partial}{\partial r}\left(\Phi_{1}+h_{1}\right)\,,
\label{euler1}
\end{equation}
\begin{equation}
\frac{\partial v_{\varphi 1}}{\partial t}+\Omega \frac{\partial v_{\varphi 1}}{\partial \varphi}+\frac{\kappa^{2}}{2\Omega} v_{r1}=-\frac{1}{r}\frac{\partial}{\partial \varphi}\left(\Phi_{1}+h_{1}\right)\,,
\label{euler21}
\end{equation}
where we have assumed that the disk is barotropic and the equation of state is $p=K \Sigma^{\gamma}$, where $\Sigma$ is the surface density, and $\gamma$ is the polytropic index. On the other hand $v_{r}$ and $v_{\varphi}$ are the radial and azimuthal velocity components, and $h_1$ is defined as $h_1=c_s^2 \Sigma_1/\Sigma_0$, where $c_s^2$ is the sound speed, and can be written as follows
\begin{align}
c_s=\sqrt{\frac{d p}{d\Sigma}}=K\gamma\Sigma^{\gamma-1}\,.
\label{cs}
\end{align}
Also $\kappa$ is the epicycle frequency given by
\begin{equation}
\kappa(r)=\sqrt{r \frac{d\Omega^{2}}{dr}+4 \Omega^{2}}\,,
\label{epip}
\end{equation}
and $\Omega$ is the angular velocity. Now restricting ourselves to the tight winding, or WKB-approximation, we can write the perturbations as (\citealp{binney})
\begin{equation}
\mathcal{Q}_1=\mathcal{Q}_a\, e^{i (k\,r +m\varphi- \omega t)}\,,
\end{equation}
where $\omega$ is the oscillation frequency and $k=2\pi/ \lambda$ is the wavenumber of the density wave. The WKB approximation works for very short wavelengths, i.e. $|k\,r| /m\gg1 $. In this case Eqs (\ref{cont1})-(\ref{euler21}) yield the following solutions for the coefficients
\begin{equation}
(m \Omega-\omega)\Sigma_a+k\Sigma_0 v_{ra}=0\,,
\label{conti3}
\end{equation}
\begin{equation}
v_{ra}=(m\Omega-\omega)k(\Phi_a+h_a) \Delta^{-1}\,,
\label{eulerr3}
\end{equation}
\begin{equation}
v_{\varphi a}=2i B (\omega-m\Omega)^{-1} v_{ra}\,,
\label{eulerphi3}
\end{equation}
where $\Delta$ and the Oort's constant of rotation $B$ are defined as 
\begin{equation}
\Delta=\kappa^2-(m\Omega-\omega)^2\,,
\end{equation}
\begin{equation}
B(r)=-\frac{1}{2}\Big(\Omega+\frac{d(\Omega r)}{dr}\Big)\,.
\end{equation}
Now if we find $h_a$ and $\Phi_a$ in terms of $\Sigma_a$, then Eq. (\ref{cont1}) yields the dispersion relation. Fortunately, using the definition of $h_1$, one can simply find $h_a=c_s^2 \Sigma_a/\Sigma_0$. Therefore, we only need to find the potential of a tightly wound spiral perturbation, $\Sigma_1=\Sigma_a e^{i(kr+m\varphi-\omega t)}$ which is a plane wave propagating along the radial direction, in the vicinity of a point $(r_0,\varphi_0)$. It is helpful to choose the $x$ axis to be parallel to the radial direction. Therefore restricting ourselves to axi-symmetric perturbations, we can write $\Sigma_1=\Sigma_{a} e^{i(k x -\omega t)}$.

On the other hand the linearized version of Eq. (\ref{poiss}) can be written as
\begin{eqnarray}
\nabla^2 \Phi_1 =4\pi G \Sigma_1 \delta(z)+\frac{\chi}{4}\nabla^2 \Big(\Sigma_1 \delta(z)\Big)\,,
\label{gauss1}
\end{eqnarray}
which takes the following form for our density wave
\begin{equation}
\nabla^2\Phi_1=\Big[\Big(4 \pi G -\frac{\chi\, k^2}{4} \Big)\delta(z)+\frac{\chi}{4}\frac{d^2\delta(z)}{dz^2}\Big]\Sigma_1\,,
\label{po3}
\end{equation}
Hence it is natural to expect that the solution takes the following form
\begin{equation}
\Phi_{1}(x,y,z,t)=\Phi_a~ e^{i (k x-\omega t)-|\zeta z|}\,,
\label{po4}
\end{equation}
which for $z\neq 0$ should satisfy the condition $\nabla^2\Phi_1=0$. Therefore, one may straightforwardly verify that $\zeta=k$. In order to find $\Phi_a$ with respect to $\Sigma_a$, we integrate the both sides of Eq. (\ref{po3}) along the $z$ axis in the interval $(-\xi,+\xi)$, and then let $\xi\rightarrow 0$ in the final result. So, using Eq. (\ref{po4}) and the properties of the Dirac's Delta function, we get
\begin{equation}
\Phi_a=-\Big(\frac{2\pi G}{k} -\frac{k \chi}{8}\Big)\Sigma_a\,,
\end{equation}
where, without loss of generality, we have restricted the analysis to $k>0$ which, in the density wave language, corresponds to a trailing spiral density wave. By setting $\chi$ to zero one can immediately recover the corresponding expression in the Newtonian gravity. Substituting this equation into Eq. (\ref{eulerr3}), and regarding $h_a=c_s^2 \Sigma_a/\Sigma_0$ together with Eq. \eqref{conti3}, one may easily find the following dispersion relation
\begin{equation}\label{dis_rel}
\omega^2=\kappa^2-2\pi G \Sigma_0 k+ c_s^2 k^2+\frac{\chi \Sigma_0}{8} k^3\,.
\end{equation}
It is clear that the EiBI corrections induce a cubic term to the dispersion relation. This term can substantially influence the dynamics of the perturbations. It is somehow clear that the sign of $\chi$ is a crucial factor to determine the consequences of EiBI gravity. Using Eq. \eqref{dis_rel} we expect that positive $\chi$ gives rise to stabilizing effects, and the negative $\chi$ supports the instability. In the following we carefully discuss both cases. For the sake of simplicity, let us define a dimensionless wavenumber $X$ and a dimensionless parameter $\beta$ as
\begin{align}\label{X_and_beta}
X=\frac{k}{k_{\text{crit}}}, ~~~~~~~\beta=\frac{\chi \kappa^4}{64 \pi^3 G^3 \Sigma_0^2}\,,
\end{align}
where the critical wavenumber is conveniently defined as $k_{\text{crit}}=\kappa^2/2\pi G \Sigma_0$. Naturally $\beta$ is the most important parameter in this paper, since it parametrizes the significance of EiBI gravity in the stability of the system. Using these new parameters the stability condition $\omega^2>0$ can be expressed as
\begin{equation}
Q^2>\frac{4(X-1)}{X^2}-4\beta X\,
\label{con1}
\end{equation}
where $Q$ is the so-called Toomre's stability parameter given by 
\begin{equation}
Q=\frac{\kappa c_s}{\pi G \Sigma_0}\,.
\end{equation}

\begin{figure*}
\begin{center}
\includegraphics[width=7.5cm]{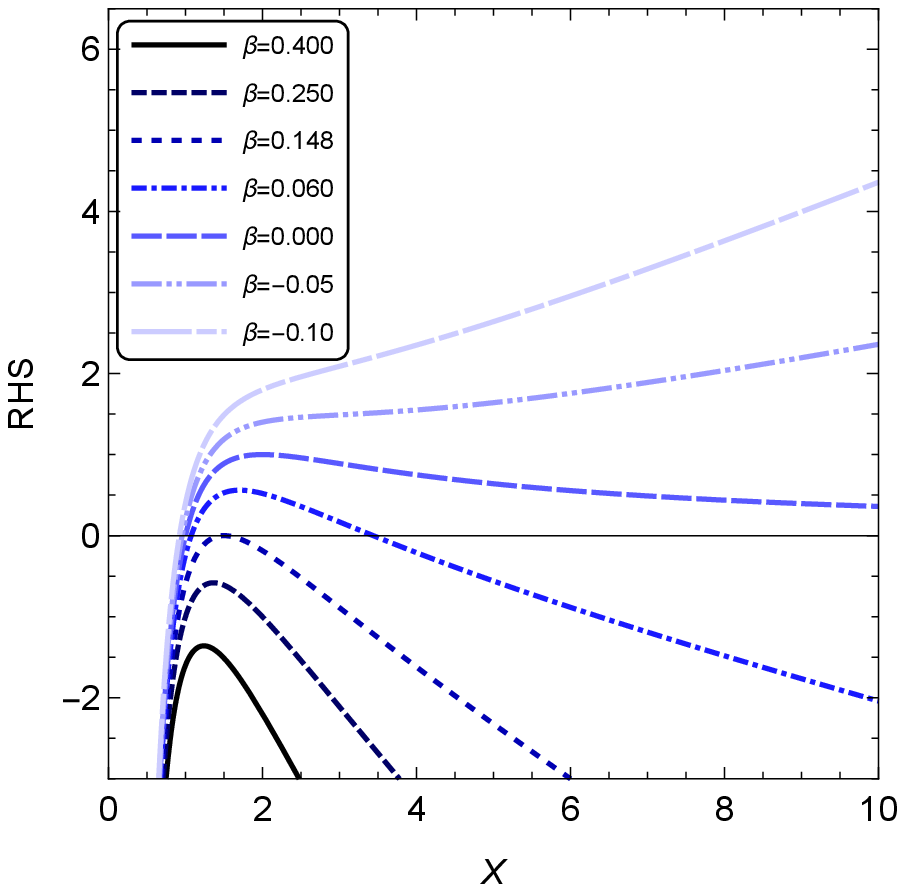}\hspace{1cm}
\includegraphics[width=7.5cm]{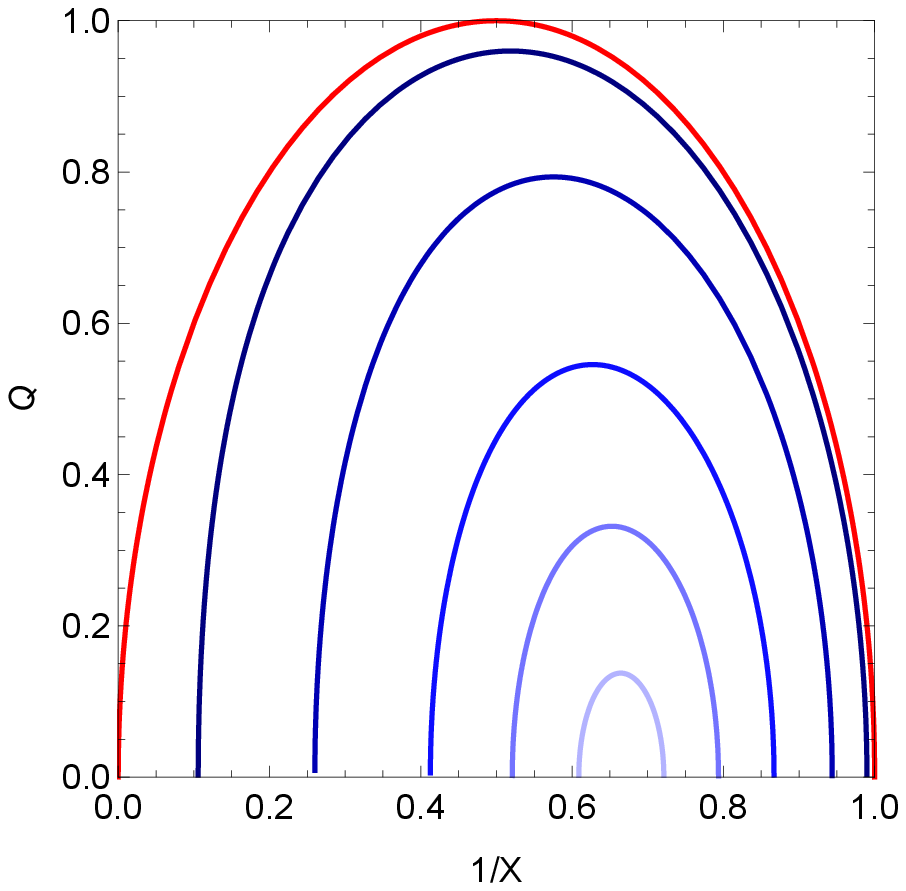}
\caption{ \textit{Left panel:}  The RHS of Eq. \eqref{con1} versus $X$ for a variety of $\beta$.
 \textit{Right panel:}  The boundary between stable and unstable axisymmetric perturbations in a fluid disk in the context of EiBI gravity. The red curve depicts the Newtonian limit. Solid blue curves show the boundary in EiBI for different values of $\beta$. From up to down the magnitude of $\beta$ increases.}
\label{FIG1}
\end{center}
\end{figure*}

It is helpful to remind that in Newtonian gravity, where the last term in the right hand side (RHS) of (\ref{con1}) is zero, the stability conditions is automatically satisfied for $X\leq 1$ since RHS of (\ref{con1}) gets negative. This means that the disk is stable against all perturbations with wavelengths larger than $\lambda_{\text{crit}}=4\pi^2 G \Sigma_0/\kappa^2$. On the other hand, if $Q>1$ then all wavelengths with $\lambda\leq \lambda_{\text{crit}}$ will be stable. However the situation in EiBI gravity is more complicated. 

The left panel of Fig. \ref{FIG1} shows the behavior of the RHS of Eq. \eqref{con1} as function of $X$, for various values of $\beta$ which encodes the EiBI gravity. The plot deserves some comments. First, there exists a critical value of $\beta$ above which
the RHS is negative for all wavenumbers and, therefore, the disk is stable. Such a value, $\beta_c=4/27\simeq 0.148$ is depicted in the figure. Second, with increasing $\beta$ the system becomes more stable. Third, for $\beta\geq0$, the RHS of Eq. \eqref{con1} has a maximum. Consequently, to find a general criterion for stabilizing all perturbations, the Toomre's parameter $Q$ should be larger than this maximum value. Finally, for $\beta<0$ or equivalently for $\chi<0$, there is no such a maximum and, surprisingly, it is never possible to find a condition to stabilize all the linear perturbations in the system. This directly means that all the rotating thin disks are locally unstable in the context of EiBI gravity with negative free parameter $\chi$. This effect is somehow expected. A negative value of the parameter $\chi$ can be translated in a negative value of the corresponding effective pressure which favor the gravitational collapse, instead of contrasting it. It turns out that such a behavior is not related to the magnitude of value of the parameter $\beta$ chosen in the figure. Nevertheless, it is necessary to mention that this is not the case for non-rotating and pressureless infinite mediums. In other words, in such systems although negative $\chi$ leads to smaller Jeans mass compared to the Newtonian case (and so has destabilizing effects), does not make the system unstable to all perturbations, for more details on the stability of non-rotating systems see \cite{idm2017a}. The unusual behavior of EiBI with negative $\chi$ has also been seen in other contexts. For example, in cosmological context, $\chi<0$ drives the imaginary effective sound speed instabilities \citep{Avelino2012b}, and all the scalar, vector, and tensor modes are also unstable \citep{Yang2013}. Finally, Newtonian polytropic stars also result to be unstable \citep{Pani2011,pani}.

Now let us focus on the opposite case where $\beta$ is positive. This case is more interesting from physical and observational point of view. It is necessary to mention that, as we will discuss in the subsequent sections, in reality $\beta$ is a small parameter including the EiBI correction. For example, in the solar neighborhood $\beta$ is of the order of $10^{-35}$. Now let us investigate systems with $\beta\leq 4/27$. In this case, instead of $X\leq 1$ in Newtonian gravity, we have the following intervals for stable modes in EiBI gravity 
\begin{equation}
X\leq A~~~~~~~\text{and}~~~~~~~X\geq B\,,
\label{con2}
\end{equation}
where $A$ and $B$ are the positive roots of RHS of (\ref{con1}). Keeping in mind the smallness of $\beta$, we find the following expressions for $A$ and $B$
\begin{equation}
A\simeq 1+\beta+3 \beta^2+ O(\beta^3)\,,
\end{equation}
\begin{equation}
B\simeq -\frac{1}{2}+\frac{1}{\sqrt{\beta}}-\frac{3\sqrt{\beta}}{8}-\frac{\beta}{2}-\frac{105}{128}\beta^{\frac{3}{2}}-\frac{3}{2}\beta^2+ O(\beta^{\frac{5}{2}})\,,
\end{equation}
condition (\ref{con2}) directly means that there are wavelengths which could be unstable in Newtonian gravity, while are stable in EiBI gravity. The clearest way to see this fact is to plot the boundary between stable and unstable waves using the curve $Q=Q(Y)$ obtained from $\omega=0$, where the dimensionless wavelength $Y$ is defined as $Y=1/X$, see right panel in Fig.  \ref{FIG1}. This curve intersect the horizontal axis $Y$ in to wavelengths $Y_A=1/A$ and $Y_B=1/B$. Therefore the interval $Y_B<Y<Y_A$ is the interval which, in principle, is unstable unless we increase the Toomre's parameter to suppress the instability. The corresponding interval in Newtonian gravity is $0<Y<1$. However since $Y_B>0$ and $Y_A<1$, we conclude that this interval shortens in EiBI gravity that has stabilizing effects on the system. This point is clear from the right panel of Fig. \ref{FIG1}. 

In the right panel of Fig. \ref{FIG1}, the red curve shows the boundary in the Newtonian gravity. On the other hand, black curves are corresponding boundaries for different values of $\beta$ in EiBI gravity. From up to down we increase the magnitude of $\beta$. It is obvious that by increasing $\beta$, the instability interval shortens and the stability range gets larger. On the other hand, the most unstable mode, the first mode which gets unstable when the Toomre's parameter is decreased, shifts to right, i.e. to larger wavelengths. In the Newtonian description this wavelength is equal to $0.5 \lambda_{\text{crit}}$. However, depending on the value of $\beta$, it is larger in EiBI gravity. For example when $\beta\rightarrow 4/27$ this most unstable wavelength reaches $0.64 \lambda_{\text{crit}}$. 

More importantly, by increasing the $\beta$ parameter, the maximum value of $Q$ required for stability decreases. This fact again shows that EiBI has strong stabilizing effects. One can easily maximize the RHS of (\ref{con1}) and find an exact form for a new version of the Toomre's criterion. The result can be written as
\begin{equation}
Q>2 \Bigg(\frac{72^{\frac{1}{3}} \beta  f(\beta )-6 \beta ^2 \sqrt[3]{f(\beta )}-9 \sqrt[3]{3} \beta ^2 f(\beta )^{\frac{2}{3}}}{\left(f(\beta )^{\frac{2}{3}}-\sqrt[3]{3} \beta \right)^2}\Bigg)^{\frac{1}{2}}\,,
\label{toom1}
\end{equation}
where $f(\beta)$ is defined as
\begin{equation}
f(\beta)=\sqrt{3\beta ^3 (27 \beta +1)}+9 \beta ^2\,,
 \end{equation}
 the RHS of criterion (\ref{toom1}) is always smaller than $1$ for $\beta\leq 4/27$. Expanding it in terms of $\beta$ we find
 \begin{equation}
 Q > 1-4 \beta-64\beta^3 + O(\beta^4)\,.
 \end{equation}
 Since the RHS is smaller than one, EiBI stabilizes the disk requiring a smaller $Q$, compared with the Newtonian case, to suppress the local perturbations. However, it is clear that, in order to see a meaningful differences between Newtonian and EiBI descriptions, it is necessary to study astrophysical disks in which $\beta$ is not too small. In the subsequent sections we will study the growth rate of unstable modes in massive disks around HMNS in the context of EiBI gravity, and compare it with the standard case.
 
 Before moving on to discuss the possible effects of EiBI gravity in astrophysical systems, it is important mentioning that the stabilizing behavior of EiBI is, somehow, in harmony with the main feature of the theory to resolve the singularities. In fact, as we already mentioned this theory prevents the singularity in the early universe and in gravitational collapse of noninteracting particles. In other words, the fate of a dust collapse in this theory is a regular star rather than a singularity, see \cite{pani}. Furthermore, when $\chi>0$, the EiBI corrections to the hydrodynamic equations appear as an effective pressure that supports the stability, at least in the weak field limit, and helps to suppress local gravitational collapse. From this perspective it is natural to expect stabilizing behavior in EiBI gravity.
 
 However one needs to be very careful when interpreting the Toomre's criterion. In fact the left hand side of 
 (\ref{toom1}) is also different from the standard case. In other words in order to make a final decision on the possible effects of EiBI on the stability of disks, it is necessary to take into account both sides of (\ref{toom1}). Furthermore it is necessary to note that $\beta$ is not a constant and is a function of $r$. To decouple the effects of gravity and pressure on the stability of the system, and compare differences between EiBI and standard gravity, it is instructive to study the stability of an exponential disk as a toy model.
\begin{figure*}
\begin{center}
\includegraphics[scale=0.8]{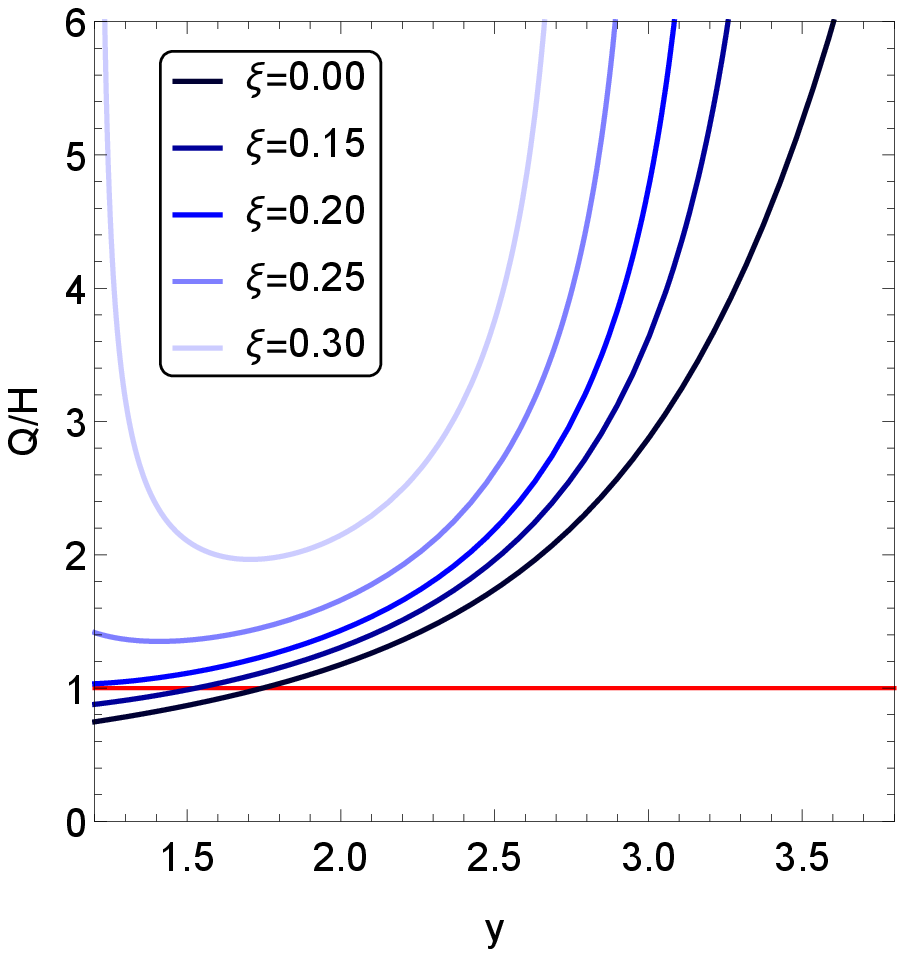}
\hspace{4mm}
\includegraphics[scale=0.85]{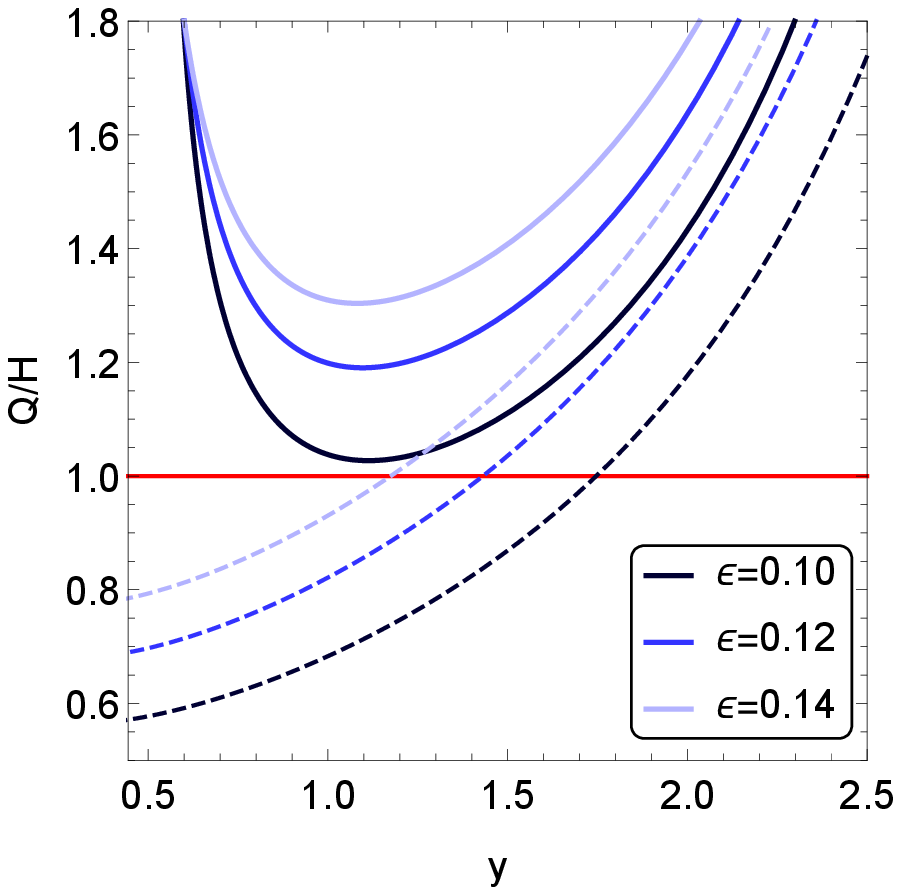}
\end{center}
\caption{The left hand side of Toomre's criterion, i.e $Q(y,\epsilon,\xi)/H(y,\epsilon,\xi)$, in terms of $y$ for EiBI and Newtonian theories. \textit{Left panel:} we fix $\epsilon=0.1$, while varying $\xi$. \textit{Right panel:} we vary $\epsilon$, while fixing $\xi=0.2$. The dashed curves are corresponding Newtonian Toomre's criterion for which $H=1$.}
\label{fig_2}
\end{figure*}
\section{Stability of exponential disks in EiBI gravity}\label{expon}
In this section we investigate the stability of an exponential disk, with the following surface density 
\begin{equation}
\Sigma(r)=\sigma_0 \exp (-\alpha r)\,.
\label{sigma_ex}
\end{equation}
where $\alpha$ is the inverse disk length scale and $\sigma_0$ is the central density. 
The first step is to find the gravitational potential of the disk in the cylindrical coordinate system $(r,\varphi,z)$. Then we will be able to find $\kappa$, $\beta$ and other relevant quantities analytically. The modified Poisson Equation (\ref{modPoiss}) in the vacuum, i.e. $z\neq 0$, coincides with the Laplace's equation. Consequently the solution is given by the cylindrical Bessel function of zero order 
\begin{equation}
\Phi_k(r,z)=\Phi_a\exp (\pm k z)J_0(k r)\,.
\end{equation}
Note that we deal with an axisymmetric disk. For $z=0$, substituting this equation into (\ref{gauss1}) and integrating both sides along the $z$ axis in the interval $(-\zeta,+\zeta)$, we arrive at
\begin{equation}
\nabla^2 \Sigma_k(r)+\frac{16 \pi G}{\chi}\Sigma_k(r)= -\frac{8 k}{\chi} J_0(k r)\,,
\end{equation}
this differential equation can be simply solved to obtain
\begin{equation}
\Sigma_k(r)=-\frac{k}{2\pi G}\Big(1-\frac{\chi k^2}{16 \pi G}\Big)^{-1}J_0 (k r)\,.
\end{equation}
A general solution for arbitrary surface density is obtained taking into account all possible values of $k$. Therefore, by defining a function $S(k)$, we can write
\begin{equation}
\Sigma(r)=-\frac{1}{2\pi G}\int_0^{\infty} k\Big(1-\frac{\chi k^2}{16 \pi G}\Big)^{-1}S(k) J_0 (k r) dk\,.
\label{hak1}
\end{equation}
In this case the gravitational potential takes the following form
\begin{equation}
\Phi(r)=\int_0^{\infty} S(k) J_0(k r) dk\,,
\label{pot2}
\end{equation}
thus, using the orthogonality of Bessel functions, one may simply find $S(k)$ from (\ref{hak1}) as
\begin{equation}
S(k)=- \frac{2 \pi G \sigma_0 \alpha}{(k^2+\alpha^2)^{3/2}}\Big(1-\frac{\chi k^2}{16 \pi G}\Big)\,,
\end{equation}
Inserting this equation into (\ref{pot2}), fortunately, we can find an analytical expression for the gravitational potential of the exponential disk in EiBI gravity as
\begin{align}\label{Phi}
\begin{split}
\Phi(y)= & \frac{2 \pi  G \sigma _0}{\alpha} \Big((\xi ^2+1) y I_1(y) K_0(y)+I_0(y) (\xi ^2 K_0(y)\\&-(\xi ^2+1) y K_1(y))\Big)\,,
\end{split}
\end{align}
where dimensionless distance $y$ is defined as $y=\alpha r/2$, and also the auxiliary variable $\xi^2=\chi\alpha^2/16\pi G$. It worth mentioning that, $\xi$ can be written in terms of EiBI characteristic wavelength $\lambda_\text{EiBI}$ as follows
\begin{align}
\xi=\left( \frac{\alpha}{2\pi}\right) \lambda_\text{EiBI}\,,
\end{align}
where EiBI wavelength is given by \citet{idm2017a}
\begin{align}
\lambda_\text{EiBI}=\sqrt{\frac{\pi |\chi|}{4 G}}\,.
\end{align}
Therefore $\xi$ is simply the ratio of the EiBI and the exponential disk's characteristic length. So, for small values of $\xi$, the EiBI wavelength gets small compared to the disk scale length, one can neglect the effects of this theory. As an example, the gravitational potential of the disk in Newtonian gravity can be recovered by setting $\xi$ to zero. On the other hand, the angular velocity is given by the following expression
\begin{equation}
\Omega^2(r)=\frac{1}{r}\Bigg(\frac{d\Phi}{d r}+c_s^2\frac{d}{dr}\ln \Sigma\Bigg)\,.
\label{omega2}
\end{equation}
Now it is straightforward to calculate the epicycle frequency and consequently the Toomre's parameter $Q$ and the function $\beta$:
\begin{align}\label{q1}
Q= &\nonumber e^{(3-\gamma)y} \sqrt{\frac{\gamma  \mu }{\eta\pi  y}} \Big[I_1(y) \Big((\xi ^2+2 (\xi ^2+1) y^2) K_0(y)\\&-(3 \xi ^2+2) y K_1(y)\Big)-I_0(y) \Big((\xi ^2+2 (\xi ^2+1) y^2) K_1(y)\\&-(5 \xi ^2+4) y K_0(y)\Big)+\frac{\gamma\mu}{2\pi\eta}e^{2y(1-\gamma)}\Big(2y(\gamma-1)-3\Big)\Big]^{1/2}\,,\nonumber
\end{align}
where the dimensionless stability parameters $\eta$ and $\mu$ are defined as 
\begin{equation}
\eta=\frac{\sigma_0 G }{\alpha c^2}, ~~~~~~ \mu=\frac{K \sigma_0^{\gamma-1}}{c^2}\,,
\label{etamu}
\end{equation}
and the function $\beta$ in terms of $y$ and $\xi$ is given by
\begin{align}\label{beta}
\beta= & \frac{\xi ^2 e^{-4 (\gamma -1) y}}{16 \pi ^2  y^2}\frac{\mu^2}{\eta ^2}\Big[
\gamma   e^{2 y} (2 (\gamma -1) y-3)+2 \pi  \frac{\eta}{\mu}  e^{2 \gamma  y}\\\nonumber&\Big(
I_1(y) \Big[\left(\xi ^2+2 \left(\xi ^2+1\right) y^2\right) K_0(y)-\left(3 \xi ^2+2\right) y  K_1(y)\Big]\\\nonumber
& +I_0(y) \Big[\left(5 \xi ^2+4\right) y K_0(y)-\left(\xi ^2+2 \left(\xi ^2+1\right) y^2\right)
   K_1(y)\Big]
\Big)
\Big]^2.
\end{align}

It is important mentioning that we call $\eta$ and $\mu$ as stability parameters in the sense that they directly control the occurrence of gravitational instability in the system. More specifically, $\eta$ can be regarded as a representative for gravity, and similarly it is clear from the definition that $\mu$ is related to the pressure's role in the system. In fact using Eq. (\ref{q1}) one may easily show that for our exponential disk, the $Q$ parameter is a function of $\mu/\eta$. Therefore, at least at $y>3/2(\gamma-1)$, increasing $\mu/\eta$ makes the $Q$ parameter larger. Albeit this does not necessarily mean that the system gets more stable against local perturbations. One should note that since $\beta$ is a function of $\mu/\eta$ (see Eq. (\ref{beta})), the right hand side of stability criterion (\ref{toom1}) is also a function of $\mu/\eta$. Consequently, here we deal with a stability issue only with two parameters $\xi$ and $\epsilon$, where $\epsilon$ is defined as
\begin{equation}
\epsilon=\frac{\mu}{\eta}=\frac{\alpha K \sigma_0^{\gamma-2}}{G}\,.
\end{equation}
and as we already mentioned $\xi$ is related to the free parameter of the theory. Before moving on to discuss the stability criterion (\ref{toom1}) in terms of $\xi$ and $\epsilon$, let us mention two restrictions on the stability parameter $\epsilon$. It is necessary for $\Omega$ and also $\kappa$ to be real quantities. This imposes two restrictions on $\epsilon$ at each radius. Considering Eqs \eqref{cs}, \eqref{sigma_ex}, \eqref{Phi}, and \eqref{omega2}, one can simply calculate $\Omega$ and $\kappa$. Then the mentioned conditions are given by
\begin{align}\label{cons1}
 \epsilon\leq &\nonumber \pi e^{2(\gamma-1)y}\gamma^{-1}  \Big[I_0(y) \Big(2 \left(\xi ^2+1\right) y K_0(y)-\xi ^2 K_1(y)\Big) +I_1(y) \\& \Big(\xi
   ^2 K_0(y)-2 \left(\xi ^2+1\right) y K_1(y)\Big)\Big]
\end{align}
and also
\begin{align}\label{cons2}
  \epsilon (3-2 &\nonumber(\gamma -1) y)\leq  2 \pi   e^{2 (\gamma-1)  y} \gamma^{-1} \Big[ I_1(y) \Big(\left(\xi ^2+2 \left(\xi ^2+1\right) y^2\right)
  \\\nonumber& K_0(y)-\left(3 \xi ^2+2\right)
   \times y K_1(y)\Big) +I_0(y) \Big(\left(5 \xi ^2+4\right) y
   K_0(y)\\
   &-\big(\xi ^2+2 \left(\xi ^2+1\right) \times y^2\big) K_1(y)\Big)\Big]\,.
\end{align}

Note that, the parameters that will be used hereafter, are inside the realm of the validity of these constraints. Now, let us write the stability criterion (\ref{toom1}) as $Q>H$. It should be noted that the original version of this criterion is $Q^2>H^2$. Consequently the system is stable in radii where $H$ is imaginary, i.e. $H(y,\epsilon,\xi)\in\mathbb{I}$. On the other hand we recall that if $\beta(y,\epsilon,\xi)>\beta_c$ then the disk is stable against all perturbations at $y$. Consequently if $\beta(y,\epsilon,\xi)<\beta_c$ then the stability criterion can be written as
\begin{equation}
\frac{Q(y,\epsilon,\xi)}{ H(y,\epsilon,\xi)}>1\,.
\label{TC}
\end{equation}
As mentioned before, in the Newtonian case we have $\xi=0$ and $H=1$. Whenever the left hand side of Eq. \eqref{TC} is less than unity, the system will be unstable there. Now we are ready to plot the modified Toomre's criterion and compare both theories. 
Results have been shown in Fig. \ref{fig_2}. The left panel shows the response of the system for different values of $\xi$. The black line for which $\xi=0$, belongs to Newtonian case, and it is clear that the disk is unstable at small radii. On the other hand by increasing the dimensionless free parameter $\xi$, the disk gets more stable. Therefore one may straightforwardly conclude that EiBI has stabilizing effects.  On the other hand, it is clear that the deviation between two theories can be considerable.

On the other hand, the right panel shows the stability criterion for different values of $\epsilon$ in both theories. More specifically, the solid and dashed curves depict EiBI and Newtonian gravity respectively. Note that these curves begin from radius after which the conditions \eqref{con1} and \eqref{con2} are satisfied in EiBI. We see that increasing the stability parameter, $\epsilon$ causes more stability in both theories. This means that increasing the pressure support stabilizes the disk in both viewpoints.  

\subsection{Growth rate of axisymmetric perturbations}
For a closer examination of the gravitational stability of the disk, let us consider the growth rate of the unstable modes. The dispersion relation 
in Eq.  \eqref{dis_rel} can be rewritten in the form
\begin{align}
s^2=-1-q^2+\frac{2\pi G\Sigma_0 q}{c_s \kappa}-\frac{\chi\Sigma_0\kappa q^3}{8 c_s^3}\,,
\label{s2}
\end{align}
where we have defined
\begin{align}
& s=\frac{i\omega}{\kappa}\,, \quad \text{and} \quad q=\frac{c_s k}{\kappa}\,.
\end{align}
Using Eqs  \eqref{cs}, \eqref{sigma_ex}, and \eqref{etamu}, Eq. \eqref{s2} can be expressed as follows
\begin{figure*}
\centering
\includegraphics[scale=0.9]{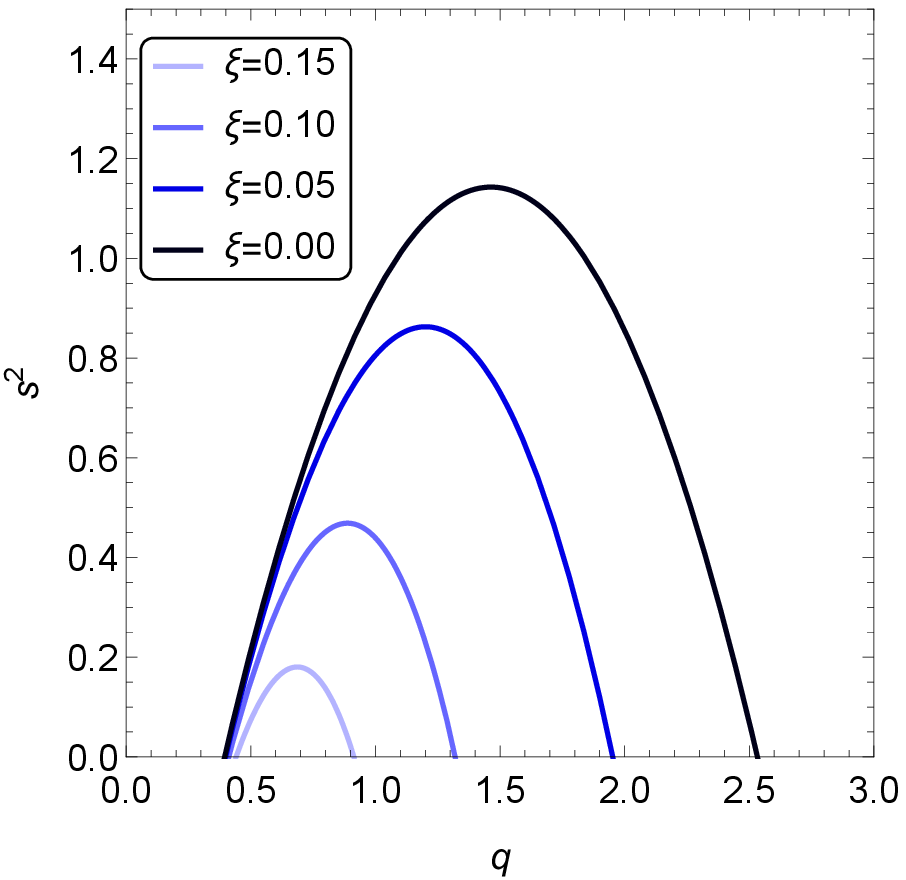}\hspace{5mm}
\includegraphics[scale=0.9]{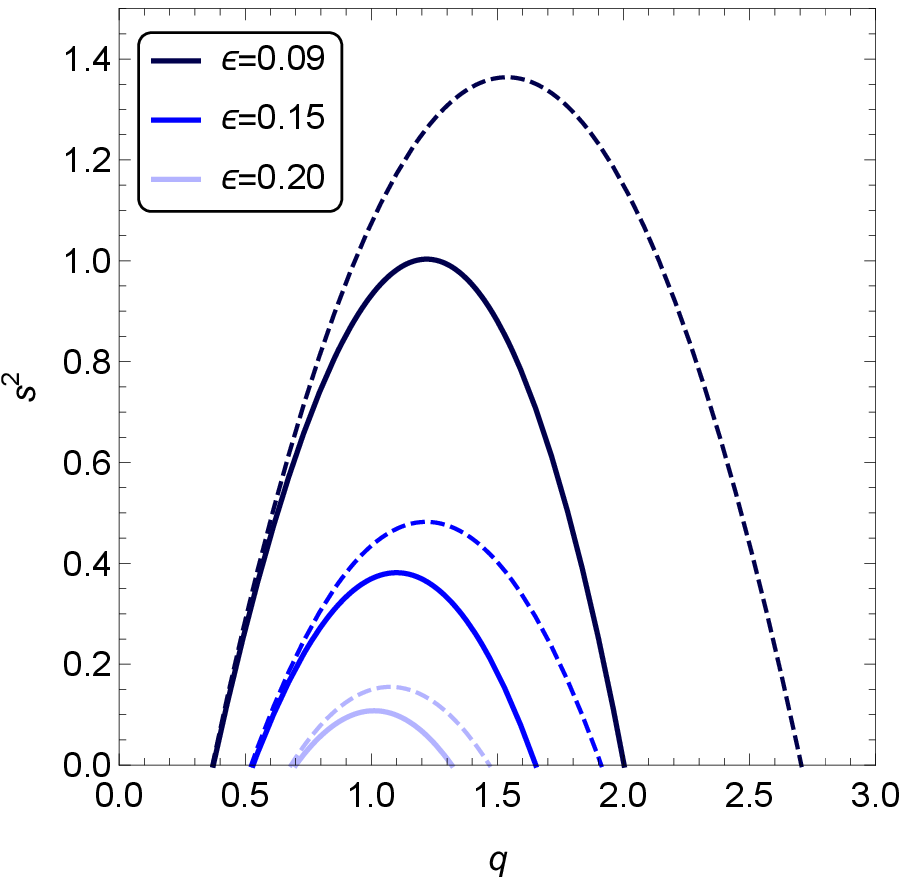}
\caption{The squared dimensionless growth rate $s^2$ with respect to the dimensionless wavenumber $q$. \textit{Left panel:} in this panel $\epsilon$ is fixed to $0.1$, and $s^2$ is plotted for different values of $\xi$. \textit{Right panel:} in this panel $\xi=0.05$ and stability parameter $\epsilon$ takes different values. Dashed curves belong to Newtonian gravity. Both panels have been plotted at $y=1$.}
\label{Rates}
\end{figure*}
\begin{align}\label{rate}
s^2  = -1-q^2&+\left(\sqrt{\frac{8y}{\gamma\epsilon}} \pi e^{y(-3+2\gamma)}\right)q\varsigma^{-1/2}-\\&\nonumber\Bigg(\sqrt{\frac{2}{y}}\frac{e^{y(-5+2\gamma)}\pi \xi^2}{\gamma^{3/2}\epsilon^{3/2}}\Bigg) q^3\varsigma^{1/2}
\end{align}
where, for the sake of simplicity, the following quantity has been defined
\begin{align}
\varsigma=& e^{2y}\left( -3+2y(-1+\gamma)\right) \gamma\epsilon+2e^{2y\gamma}\pi\Big[I_1(y)\\&\nonumber\Big(\left[  \xi^2+2y^2(1+\xi^2)\right] K_0(y)-y(2+3\xi^2)K_1(y)\Big)\\\nonumber
&+I_0(y)\left[ y(4+5\xi^2)K_0(y)-(\xi^2+2y^2(1+\xi^2))K_1(y)\right] \Big]
\end{align}

The growth rate is shown in Fig. \ref{Rates}. In both panels, without loss of generality, we fixed  $y=1$.

In the left panel  we depict Eq. \eqref{rate} for various values of $\xi$ while fixing $\epsilon=0.1$. It turns out that the overall behavior does not change for other allowed values of $\epsilon$.
The black curve represents the Newtonian dispersion relation. 
By increasing $\xi$, the growth rate of the growing modes decreases. This is not surprising in the sense that we have already seen the same behavior in Fig. \ref{fig_2}. In fact, keeping in mind the definition of $\xi$, and its direct dependence on the EiBI free parameter ($\chi$), one can expect that, by increasing this parameter, the effectiveness of the new term in the hydrodynamics equations (the last term in Eq. \eqref{poiss}), get stronger. As mentioned before, this term can be regarded as an effective pressure, therefore, the stabilizing role of EiBI is expected from this point of view. 

In the right panel of Fig. \ref{Rates}, we show the effect of stability parameter $\epsilon$ when $\xi$ is fixed at $\xi=0.05$. The dashed curves correspond to the Newtonian case. It is clear that both theories respond in a similar way to an increase in $\epsilon$. More specifically, larger $\epsilon$ leads to smaller growth rate. This result also is completely consistent with previous conclusions inferred from Fig. \ref{fig_2}. In other words, in both theories, i.e. Newtonian gravity and EiBI with $\chi>0$, with increasing the pressure budget of fluid, the system gets more stable against local fragmentation.

As the final remark, Fig. \ref{Rates} shows that, not only the growth rate is higher in Newtonian gravity, also the instability wavelength interval is wider compared with EiBI gravity.

 \begin{table}
 \centering
  \caption{The stability parameter $\epsilon$ for five different HMNS numeric models 
  (for the details of HMNS models see in \citealp{Hanauske2017}). Each model,  has been constructed for two different cases: high mass ($M = 1.35 \text{M}_\odot$) and low mass ($M = 1.25 \text{M}_\odot$) binaries.}
  \label{table1}
  \begin{tabular}{lcccc}
  \hline
  Model            &   $\alpha^{-1} \text{km} $  &  $\Sigma_0 \left[10^{22} \right]\frac{\text{kg}}{\text{m}^{2}}$  & $K_m \left[10^{-8} \right]\frac{\text{m}^3}{\text{kg}\text{s}^{2}}$  & $\epsilon$ \\
  \hline
  GNH3-M125   & 4.105  &  4.8  & 1.4 & 0.048  \\\vspace*{1mm} 
           
  GNH3-M135   & 3.432  &  7.3  & 1.98 & 0.086 \\
  H4-M125     & 4.897  &  3.45 & 9.73 & 0.032  \\\vspace*{1mm} 
  H4-M135     & 3.639  &  6.54 & 1.76 & 0.074  \\
  ALF2-M125   & 4.726  &  3.68 & 1.04 & 0.031  \\\vspace*{1mm} 
  ALF2-M135   & 2.680  &  11.97& 3.25 & 0.181  \\
  SLy-M125    & 2.847  &  9.83 & 2.88 & 0.150   \\\vspace*{1mm} 
  SLy-M135    & 2.911  &  1.02 & 2.75 & 0.142  \\
  APR4-M125   & 3.078  &  8.42  & 2.46 & 0.120 \\
  APR4-M135   & 3.417  &  7.40  & 2.00 & 0.086 \\                                                
 
  \hline
  \end{tabular}
  \end{table}
\begin{figure*}
\centering
\includegraphics[scale=0.9]{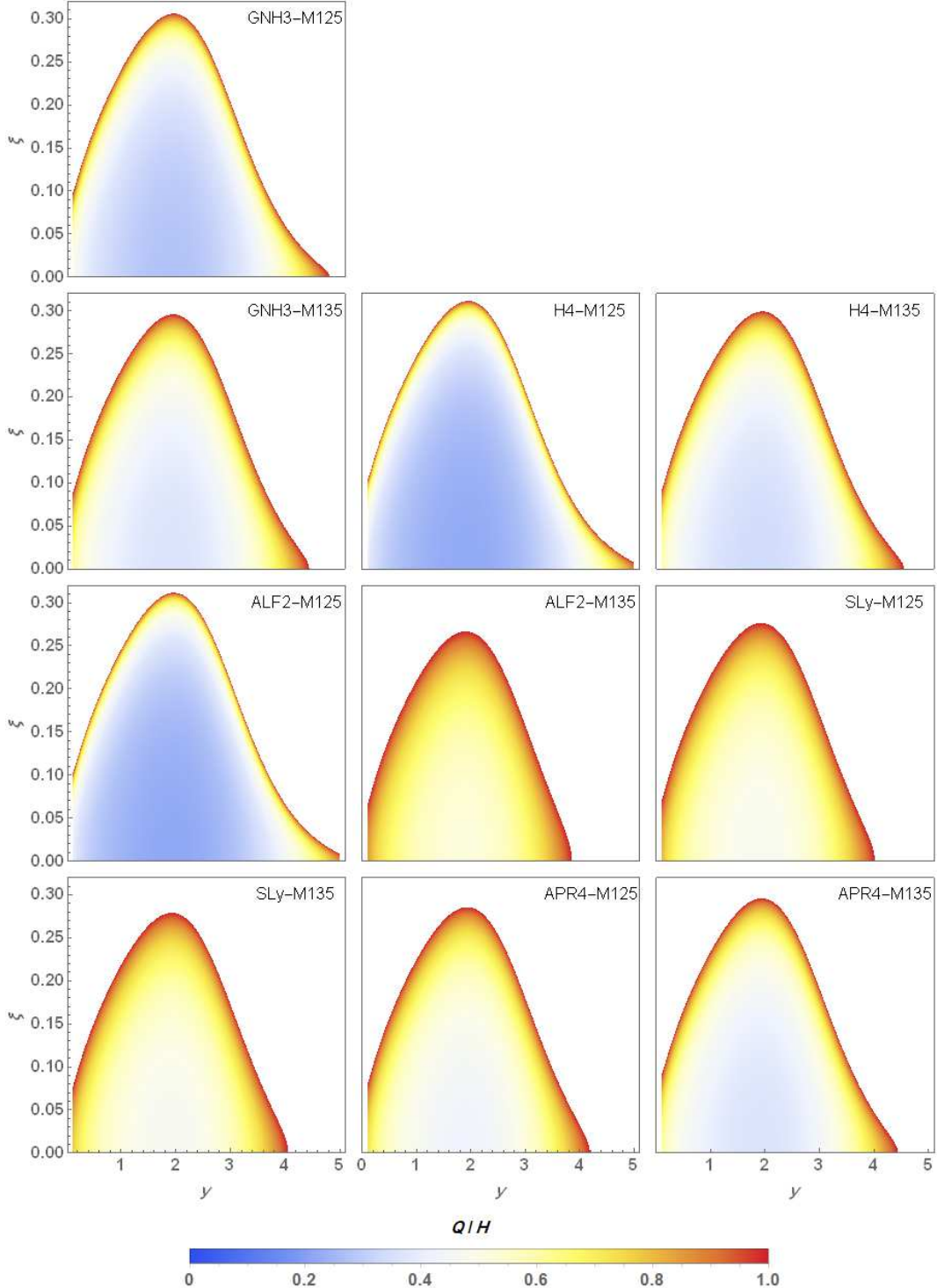}
\caption{The modified Toomre's criterion, density plot in terms of $\xi$ and dimensionless radius $y$ for models summarized in Table \ref{table1}. Each panel has been titled with the corresponding model. The white era in all models indicates the stability zone.\label{dp}}
\end{figure*}


\section{an astrophysical application}\label{applications}
Now let us use our toy model to crudely investigate the local stability of disks around HMNS. A HMNS can be formed from the remnant of a neutron star binary merger. At the moment, these binaries are at the forefront of astrophysical studies because of their importance in detection of the gravitational waves \citep{Abbott2017}. 
Here we want to apply our results to the differentially rotating and relativistic disks around HMNSs. Of course a relativistic description is required to investigate such a system. For example see \citet{kazemi} for relativistic correction to the Toomre's criterion at the first post-Newtonian approximation. However we are interested on the general behavior of EiBI gravity (which deviates from standard gravity only in high density regime). Consequently our analysis in this section can be considered as an estimation procedure in order to reveal some main features of EiBI gravity. For more complete studies it is necessary to include relativistic corrections properly.

According to  \cite{Ellis2018}, the power spectrum of the gravitational radiation from a neutron star binary merger, can be significantly affected by the existence of fragmented matter inside the disk. So, it can be interesting to search for such fragmentation by means of our toy model. We mention that this system is prone to several instabilities like Kelvin-Helmholtz and the dynamical bar instability \cite{Shibata2000}. This paper shows that the system can be unstable to formation of a twofold symmetric bar. Moreover, the magnetorotational instability (MRI) can arise and substantially influence the magnetic field distribution in the star (\citealp{Siegel2013}; \citealp{Duez2006}).

On the other hand, the fully analytic description of the neutron star merging is almost impossible. Therefore the physical properties of this system have been widely investigated using fully general relativistic simulations, for instance see \cite{Hanauske2017}. Beside this technical difficulties, it should also be mentioned that internal structure of the neutron stars is still too far from being totally understood.

Keeping in mind, all the mentioned difficulties, as a crude estimation, we find the values of $\eta$ and $\mu$ (and consequently $\epsilon$) for different models studied in \cite{Hanauske2017}. This task can be done by considering  definitions of these parameters, and estimating $\alpha$, $\Sigma_0$, $K$, and $\gamma$ from the simulations. Then we investigate the local stability of them in the context of EiBI gravity.

One of the convenient equation of states (EOSs) for neutron stars is the piecewise polytropic EOS  which, in addition to the cold part, contains a thermal part \citep{Read2009}. Here, in agreement with what is shown in \cite{Hanauske2017} for the disk area, we have ignored the thermal part. Finally we simply set the EOS of the disk as $p=K \Sigma^{\gamma}$.

To find $\Sigma_0$ and $\alpha$, although the mentioned models in \cite{Hanauske2017} have a central core, we assume a smooth and exponential profile for the density. This assumption is consistent with the numeric profiles presented in \cite{Hanauske2017}. Using this approximation we assume that integration of $\Sigma$ over the disk surface, from zero to the disk inner radius (the radius where disk starts at) gives the mass of the core.  On the other hand, integration from the disk inner radius to the outer radius (we set it to 25 km) gives the disk mass. In this case remembering that $\Sigma(r)=\Sigma_0 e^{-\alpha r}$, and $y=\alpha r/2$, one can simply find $\alpha$ and $\Sigma_0$ for the given models.

Following \citet{Siegel2013}, the values of $\gamma$ and $K$ are fixed to $2$, and $0.014 ~ \text{m}^5 \text{kg}^{-1} \text{s}^{-2}$ respectively. However, in our toy model we deal with a thin disk. Therefore $K$ should be estimated to be valid for our two dimensional model. For this task, considering a small thickness ($\delta$) for the disk, we can estimate $K$ for our model as
\begin{align}
K\simeq0.014 \,\delta^{-2}\,.
\end{align} 
To have a better estimation, we consider three thicknesses: $\delta=[1/3, 1/4, 1/5]\alpha^{-1}$; and then we take the mean value of them, i.e., $K_{\text{m}}$. Now it is straightforward to estimate the stability parameter $\epsilon$ for each model. The result are shown in Table \ref{table1}. 
It is worth mentioning that, considering the constraints \eqref{cons1}, \eqref{cons2}, the allowed range of $y$ gets limited. For example, for $\epsilon=0.048$, and $\xi=0.1$, the allowed radii should satisfy $y>0.045$. This condition for $\xi=0.3$ and $\epsilon=0.048$, is $y>0.168$. One can see that, increasing $\xi$, shorten the range of $y$ where satisfy the constraints.

For each model, the left hand side of the stability criterion \eqref{TC}, is plotted as a density plot in terms of $\xi$ and $y$ in Fig. \ref{dp}. The white area in the panels indicates the stable regions in the plane $(y,\xi)$. Therefore the horizontal axis ($\xi=0$) shows the stability of the disk in Newtonian description. This plot reveals an interesting behavior of the disk in EiBI. It is clear that in the Newtonian case all the disks are unstable in the interval $y\lesssim 4$. This means that at outer radii $y\gtrsim 4$, as expected, all the disks are stable against local fragmentation.

However the situation in EiBI is totally different. It is clear that by increasing $\xi$ the instability interval becomes shorter. In other words, as we have already mentioned EiBI has stabilizing effects. It is also seen that when $\xi$ is large enough, then disk gets stable at small radii. Consequently, in this case, the disk is unstable only on an annulus. As one can easily infer from Fig. \ref{dp} the width of the annulus gets narrower with increasing $\xi$. It is seen that if we increase the EiBI free parameter $\chi$ is such a way that $\xi>0.31$, then all the disk models get stable. It turns out that this condition coincides with $\beta>\beta_c$, which can be written in terms of $\xi$ as follows
\begin{equation}
\xi> \sqrt{\frac{4}{27}}\frac{2 \pi G \alpha \Sigma_0}{\kappa^2}\,,
\end{equation}
in fact if in the allowed radial range, $\xi$ is larger than the maximum value of RHS of this equation, then the disk gets stable everywhere. Interestingly this condition can be rewritten as 
\begin{equation}
\lambda_\text{EiBI}>\sqrt{\frac{4}{27}}\lambda_{\text{crit}}\,,
\end{equation}
where the critical wavelength is defined as $\lambda_{\text{crit}}=2\pi/k_{\text{crit}}=4\pi ^2 G \Sigma_0/\kappa^2$. It is interesting to mention that the characteristic wavelength of EiBI gravity appears as an important length scale in our stability analysis.
In other words, the magnitude of $\chi$ (or equivalently $\xi$) is a crucial factor to determine the stability. Therefore, let us determine the allowed range for $\xi$. As we already mentioned $\chi$ needs to be smaller than $10^{-3} \text{kg}^{-1}\text{m}^5 \text{s}^{-2}$. Thus using the definition of $\xi$ we can write
\begin{align}
\xi\lesssim 5.5\times 10^2 \alpha
\end{align}
On the other hand, our calculations show tat  $\alpha^{-1}\simeq 3.5 \times 10^3 ~ \text{m}$. Consequently, the allowed range for $\xi$ is written as $\xi \lesssim 0.16$. This means that, in HMNS system, EiBI gravity can not totally suppress the instability. However, there are still some serious deviations between EiBI and standard gravity. 

\section{discussions and conclusions}\label{conc}
In this paper, we have investigated fragmentation in a (fluid) thin disk, in the context of Eddington inspired Born-Infeld gravity. More specifically we have found a modified version of the Toomre's criterion to be valid in EiBI gravity. To do so, we first reviewed the weak field limit of the theory. Then using the first order perturbative analysis, and assuming the WKB approximation, we found a dispersion relation for the propagation of the density waves on the surface of the self-gravitating disk. We explicitly showed that when the only free parameter of the theory is negative ($\chi<0$) then the disk is unstable to short wavelengths and it is not possible to prevent the instability. It should be emphasized for short wavelengths the WKB approximation works perfectly. Therefore our claim for the existence of instability is reliable.

On the other hand we showed that when $\chi>0$ then the system can be locally stabilized if a Toomre like criterion is satisfied. In fact, the radius $y$ is stable against local perturbations if  $\beta(y)>4/27$. This condition can be expressed with respect to the EiBI wavelength as $\lambda_{\text{EiBI}}>\sqrt{4/27}\lambda_{\text{crit}}$. If this condition does not hold, then we should check the modified Toomre's criterion derived as $Q>H$. Investigating the boundary of stability in the ($\xi$, $1/X$) plane, and also by calculating the growth rate of the perturbations, we found that EiBI with $\chi>0$ has stabilizing effects. This feature is reminiscent of the main feature of this theory for preventing the singularities. 

Moreover in order to apply our analysis to HMNS system, we first study the stability of an exponential toy model. We solved, analytically, the gravitational potential for the disk in EiBI gravity and applied the stability criterion. Furthermore an stability parameter $\epsilon$ has been defined to measure the relative significance of pressure against gravitation in the system. In this case the stability of the disk is a two parameter study ($\xi$, $\epsilon$). We showed that in both Newtonian and EiBI gravity the disk gets more stable by increasing $\epsilon$ and $\xi$. Finally we applied the exponential toy model to disks around simulated HMNS systems with different masses for neutron stars. The matter density in this system is high enough to expect deviations between Newtonian and EiBI gravity. Our results show that although Newtonian description predicts that the disks are unstable almost on all radii, EiBI stabilizes most parts of the disk. More specifically, depending on the magnitude of the free parameter $\chi$, only an annulus remains unstable on this disk.

\section*{acknowledgment}
This work is supported by Ferdowsi University of Mashhad under Grant NO. 46770 (08/03/1397). I.D.M  acknowledges  financial  supports from the Basque Government through the research project IT-956-16. This article is
based upon work from COST Action CA1511 Cosmology and Astrophysics Network for Theoretical Advances and Training Actions (CANTATA), supported by COST (European Cooperation in Science and Technology).







\bsp	
\label{lastpage}
\end{document}